\documentclass[10pt,aps,prl,twocolumn,showpacs,showkeys]{revtex4-1}
\usepackage{graphicx}
\usepackage{dcolumn}
\usepackage{bm}
\usepackage{color}

\begin{document}
\title{Voltage-driven v.s. Current-driven Spin Torque in Anisotropic Tunneling Junctions}
\author{Aurelien Manchon}
\affiliation{$^{1}$Division of Physical Science and Engineering, Materials Science and Eng., KAUST, Thuwal 23955-6900, Saudi Arabia.}

\begin{abstract}
Non-equilibrium spin transport in a magnetic tunnel junction comprising a single magnetic layer in the presence of interfacial spin-orbit interaction (SOI) is studied theoretically. The interfacial SOI generates a spin torque of the form ${\bf T}=T_{||}{\bf M}\times({\bf z}\times{\bf M})+T_{\bot}{\bf z}\times{\bf M}$, even in the absence of an external spin polarizer. For thick and large tunnel barriers, the torque reduces to the perpendicular component, $T_{\bot}$, which can be electrically tuned by applying a voltage across the insulator. In the limit of thin and low tunnel barriers, the in-plane torque $T_{||}$ emerges, proportional to the tunneling current density. Experimental implications on magnetic devices are discussed.
\end{abstract}

\maketitle

\section{Introduction}

Current-driven magnetization dynamics (switching and excitations) using Spin Transfer Torque\cite{slonc,tsoi} (STT) has essentially been observed in inhomogeneous magnetic structures such as spin-valves, magnetic tunnel junctions \cite{review} and magnetic domain walls \cite{reviewdw}. Indeed, in such structures an external polarizer or, equivalently, an inhomogeneous spin texture is needed to allow for angular momentum transfer from the spin momentum of itinerant carriers to the localized magnetic moments. Experimentally, this implies either complex magnetization dynamics (complex domain wall motion and transformations \cite{reviewdw}) or advanced design involving multilayered structures such as synthetic antiferromagnets \cite{saf} to stabilize the external polarizer. Therefore, controlling the magnetization direction of a single ferromagnet without the need of an external polarizer or magnetic texture would provide significant advantage for novel spin devices.\par

Among the different methods available, magnetoelectric effect \cite{meeb}, multiferroic \cite{multiferroics} and magnetostrictive \cite{magnetostriction} materials have been proposed. In Ga(Mn,As), electrical control of magnetic anisotropy through the voltage dependence of hole density has been demonstrated \cite{gamnas} and in rare-earth/transition metals compounds laser-induced magnetization reversal through inverse Faraday-effect has been achieved \cite{limr}. Besides all these mechanisms, the electrical control of interfacial magnetic anisotropy offers an elegant and promising solution \cite{gamble,nano}.\par

Indeed, recent experiments have demonstrated the possibility to manipulate the interfacial magnetic anisotropy of a thin ferromagnet by applying a bias voltage through an adjacent insulator \cite{gamble,nano}. Voltage-driven Magneto Crystalline Anisotropy (MCA) change has been observed for large bias voltages \cite{nano}. This observation has been explained in terms of voltage-controlled filling of electron orbitals at the interface between the insulator and the ferromagnet \cite{tsymbal}. In this configuration, the bias voltage generates a non-equilibrium Perpendicular Magnetic Anisotropy (PMA) that can in principle switch the magnetization direction from in-plane to out-of-plane and vice-versa. Notice that this voltage-driven PMA acts as an effective field that can not excite self-sustained magnetization precessions.\par 

Alternatively, a recent prediction \cite{manchon} suggested that the spin-orbit interaction (SOI) present in a asymmetrically designed thin single ferromagnetic layer could be used to generate current-driven effective magnetic field and to manipulate the local magnetization. This prediction has been experimentally confirmed recently \cite{miron,leonid} and followed by a number of theoretical investigations \cite{soitorque}. Similarly to the voltage-driven PMA mentioned above, one of the drawbacks of the SOI-induced torque (SOI-ST) derived in Ref. \cite{manchon} is that the torque acts as an effective magnetic field \cite{miron,leonid} that does not compete with the Gilbert damping (contrary to STT) so that no self-sustained precessions can be excited.\par
\begin{figure}[h!]
	\centering\begin{tabular}{c}
	\includegraphics[width=6cm]{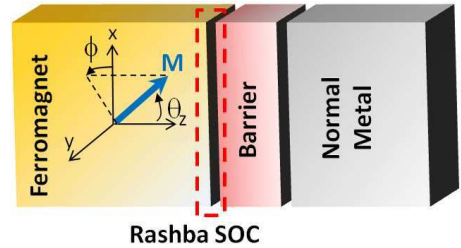}\\
	\end{tabular}
	\caption{\label{fig:Fig1}(Color online) Schematics of a Semi-Magnetic Tunnel Junction comprising a single ferromagnet and a Rashba-type SOI at the interface between the ferromagnet and the insulator.}
\end{figure}

In the present paper, we study the non-equilibrium spin transport in a Semi-Magnetic Tunnel Junction (SMTJ), comprising a single ferromagnetic layer as well as interfacial SOI at the interface between the ferromagnet and the tunnel barrier. At equilibrium (zero bias), the interfacial spin-orbit coupling is responsible for a large PMA. In the case of thick and large barrier, when tunneling conductance vanishes, the amplitude of PMA can be tuned by a gate voltage, yielding {\em voltage-controlled} PMA, as experimentally demonstrated in Ref. \cite{nano}. In the non-equilibrium regime and thin barrier limit, the interfacial SOI generates an additional {\em current-driven} spin torque on the ferromagnetic layer that can induce either magnetization switching or self-sustained magnetic precessions \cite{manchon2}. \par

\begin{figure}[h!]
	\centering\begin{tabular}{c}
	\includegraphics[width=6cm]{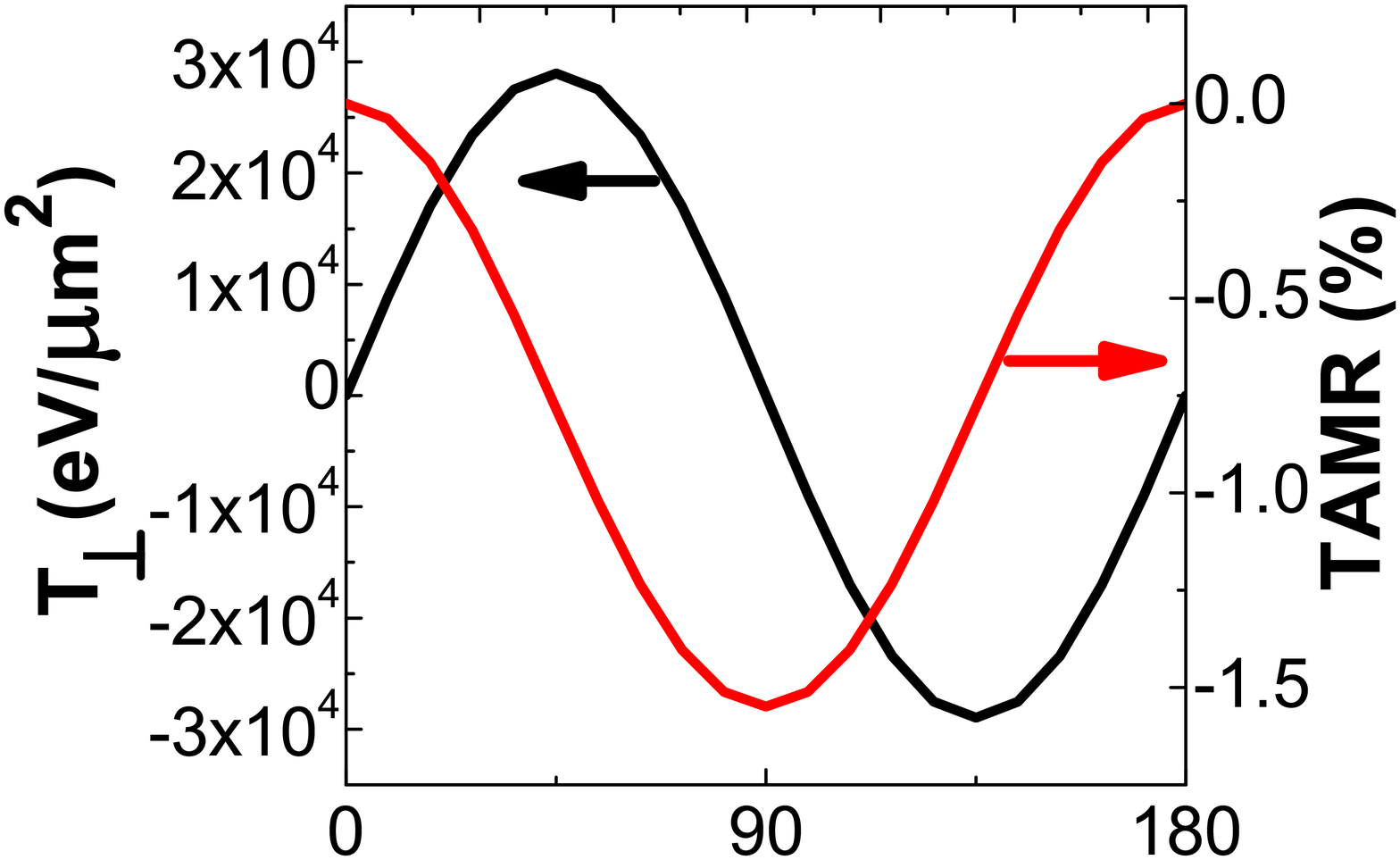}
	\end{tabular}
	\caption{\label{fig:Fig2}(Color online) Angular dependence of the TAMR and perpendicular torque at zero bias for $\alpha_R=5$eV.\AA$^2$ and $d=1$nm. The other parameters are given in the text.}
\end{figure}

\section{Theoretical Model}
SMTJs are well known to display tunneling anisotropic magnetoresistance \cite{ruster,moser,park} (TAMR): the resistance of the junction depends on the orientation of the magnetization compared to the normal of the layers (angle $\theta$ in Fig. 1). This effect has been observed in a wide range of F/I interfaces, where F is a ferromagnet (Fe, Ni, Co, GaMnAs etc.) and I is an insulator (AlOx, MgO, GaAs, etc.)\cite{ruster,moser,park}. Interestingly, TAMR is interpreted as an effect of interfacial spin-orbit coupling, such as Rashba and Dresselhaus SOI \cite{tsymbal2,matos}:
\begin{eqnarray}
{\hat H}_{SOI}=[\alpha_R(\sigma_xk_y-\sigma_yk_x)\delta(z)+\alpha_D(\sigma_xk_x-\sigma_yk_y)]\delta(z),\nonumber\\\label{eq:D}
\end{eqnarray}
where the first (second) term is the Rashba (k-linear Dresselhaus) SOI and $\alpha_{R(D)}$ is the Rashba (Dresselhaus) parameter. The former arises from structure inversion asymmetry (SIA) \cite{rashba}: when the potential drop is oriented preferentially along one direction (say, ${\bf z}$ at the interface between the insulator and the ferromagnet), it produces an electric field on the form ${\bf E}\approx - \partial_z V{\bf z}$. In the presence of finite electron velocity, this field acts on the electron spin like an effective magnetic field \cite{wrinkler}. The latter form emerges from bulk inversion asymmetry (BIA) in strained non-centrosymmetric crystals or interfaces \cite{dressel}, such as ZnS, GaAs etc. The spin-dependent energy shift associated with the spin-orbit coupling is on the form \cite{matos}:\par
\begin{eqnarray}
E_{SOI}^{\uparrow,\downarrow}({\bf k})=\pm\alpha_R(M_xk_y-M_yk_x)\pm\alpha_D(M_xk_x-M_yk_y).\nonumber\\
\end{eqnarray}
when assuming adiabatic spin dynamics (the itinerant electron spin is aligned on the local magnetization ${\bf M}$). When the magnetization is aligned on ${\bf z}$, no in-plane spin component exist and the energy shift vanishes. But when the magnetization lies in the plane of the layer, the energy shift is no more zero. As a consequence, due to the presence of interfacial SOI, the system depicted in Fig. 1 is not equivalent when the magnetization is in the plane or out-of-the plane. This gives rise to TAMR effects (see discussion in Ref. \cite{moser,matos}).\par

In the present work, we consider a SMTJ consisting in F/I/N, where F is a ferromagnetic layer, I is an insulator and N is a normal metal (see Fig. \ref{fig:Fig1}). In order to capture the most important features of the mechanism described here, we choose a minimal model only considering the most relevant materials' parameters. Matos-Abiague et al. \cite{matos} showed that the anisotropic tunneling can be satisfyingly modeled within the free electron approximation, using interfacial Rashba SOI \cite{rashba}. The authors showed that reasonable agreement with experimental data could be obtained with Rashba parameters of the order of 1-5eV.\AA$^2$ \cite{moser}. The free electron Hamiltonian of the junction reads:
\begin{eqnarray}\label{eq:H1}
{\hat H}=-\frac{\hbar^2}{2}\nabla\frac{1}{m_z}\nabla+U_z+\alpha_R{\bm \sigma}\cdot({\bf k}\times {\bf z})\delta(z).
\end{eqnarray}
Here, $\alpha_R$ is the Rashba parameter\cite{rashba} and $m_z$ is the effective mass of the electron, equal to $m_0$ in the electrodes and $m_{eff}m_0$ in the barrier. $U_z$ is the potential of the junction, given by:
\begin{eqnarray}
&&U_{z<0}=J{\bm{\hat \sigma}}\cdot{\bf M}+\frac{eV}{2},U_{z>d}=-\frac{eV}{2}\nonumber\\
&&U_{0<z<d}=U_0+\frac{eV}{2}-\frac{z}{d}eV,\nonumber
\end{eqnarray}
where $U_0$ and $d$ are the barrier height and thickness, $eV$ is the bias voltage, $J$ is the $s-d$ exchange coupling, ${\bm{\hat \sigma}}$ is the vector of Pauli spin matrices and ${\bf M}=(\sin\theta\cos\phi,\sin\theta\sin\phi,\cos\theta)$ is the magnetization direction of the ferromagnetic electrode, as depicted in Fig. \ref{fig:Fig1}. 

The present minimal model satisfyingly reproduces the TAMR results obtained by Matos-Abiague et al. \cite{moser,matos} for the case of Fe/GaAs/Au and the one obtained by Park et al. \cite{park} in (Pt/Co)$_n$/AlOx/Pt. In this article, we chose the parameters to model Fe/MgO/Au SMTJ, with $k_{Fe}^\uparrow$=1.09nm$^{-1}$, $k_{Fe}^\downarrow$=0.4nm$^{-1}$, $k_{Au}$=0.8nm$^{-1}$, $U$=1eV and $d=1$nm. As shown in Fig. 2, the TAMR displays an angular dependence in $\cos2\theta$ as already shown in Ref. \cite{moser,matos}.\par

As expressed in Eq. (\ref{eq:D}), the interfacial SOI induces an angular momentum transfer between the spin momentum and the in-plane orbital momentum of itinerant electrons. This mechanism is at the basis of our study. To the first order, this transfer is antisymmetric in ${\bf k}$ and cancels out (${\hat H}_R(-{\bf k})=-{\hat H}_R({\bf k})$). However, due to the presence of the local $s-d$ exchange, at the second order in SOI, the transfer is not zero, yielding an additional spin density in the plane of the layers [$\sigma_{x,y}$ in Eq. (\ref{eq:D})]. Interestingly this spin density can in turn exert a torque on the local magnetization. This can be seen by writing down the spin continuity equation extracted from Eq. (\ref{eq:H1}):
\begin{eqnarray}
\frac{d{\bf m}}{dt}=\frac{i\hbar}{m}{\bm \nabla}\cdot\langle{\bm{\hat \sigma}}\otimes{\bm \nabla}\rangle-\frac{2J}{\hbar}{\bf m}\times{\bf M}+\frac{1}{i\hbar}\langle[{\bm{\hat \sigma}},{\hat H}_R]\rangle.
\end{eqnarray}
The first term on the right-hand side is the spatial divergence of the spin current in the absence of SOI, the second term is the torque exerted by the itinerant spin density ${\bf m}$ on the local magnetization ${\bf M}$ and the last term is the torque arising from the angular momentum transfer driven by the interfacial SOI. In regular spin-valves, tunnel junctions or domain walls, the last term usually vanishes and the torque is directly related to the spatial variation of the spin current (an additional spin relaxation rate is usually inserted in diffusive systems) \cite{review}. But in the present case, the SOI acts like a source/sink for transverse itinerant spin density that competes with the spin current. Since the torque is by definition perpendicular to the local magnetization ($|{\bf M}|=1$), it has the general form:
\begin{eqnarray}\label{eq:b}
{\bf T}=T_{||}{\bf M}\times({\bf z}\times{\bf M})+T_{\bot}{\bf z}\times{\bf M},\end{eqnarray}
where $T_{||}$ and $T_{\bot}$ are the in-plane and perpendicular components of the torque. In the following, we calculate the local itinerant spin density ${\bf m}$ and integrate this torque on the semi-infinite electrodes in order to account for the actual torque felt by the magnetization.

\section{Results and Discussion}
\subsection{Thick Barriers: Voltage-driven Perpendicular Torque}
Let us first consider the case of thick tunnel barriers, where $d>2$nm. This is typically the experimental configuration exploited in Ref. \cite{gamble,nano}. We find that the equilibrium torque ($V_b$=0) is dominated by the perpendicular component $T_\bot$, whose angular dependence is reported on Fig. 2. The angular dependence is on the form $\propto\sin2\theta$, which reflects the symmetry of the system: the SMTJ is physically equivalent upon the transformation $\theta\rightarrow-\theta,\theta+\pi$. Due to cylindrical symmetry, the spin torque (and TAMR) does not depend on $\phi$ (not shown). This perpendicular torque can be readily identified with the PMA, which competes with the demagnetizing field and favors the perpendicular direction. Interestingly, the perpendicular torque only weakly depends on (i) the bias voltage (the conductance is vanishingly small) and (ii) the barrier thickness (not shown - calculated deviations are less than 0.1\%).\par

\begin{figure}[h!]
	\centering
		\includegraphics[width=8cm]{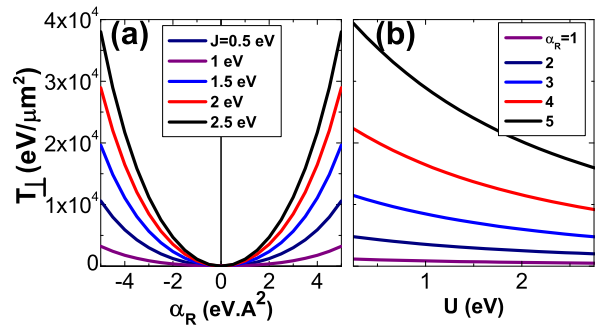}
	\caption{\label{fig:Fig3}(Color online) (a) Influence of Rashba parameter $\alpha_R$ on the equilibrium perpendicular torque $T_\bot$ for different exchanges $J$; (b) Influence of barrier height $U$ on the equilibrium perpendicular torque for different Rashba parameters $\alpha_R$ in eV.\AA$^2$.}
\end{figure}
Fig. 3(a) shows the dependence of the perpendicular torque at zero bias as a function of the Rashba parameter for different exchange energies $J$. Interestingly, the torque is quadratic as a function of $\alpha_R$. Furthermore, the dependence as a function of the barrier height is displayed in Fig. 3(b): the perpendicular torque slowly decreases when increasing the barrier height.\par

A number of experimental studies have reported that the Rashba coefficient is linear in bias voltage \cite{nitta}: $\alpha_R=\alpha_R^0+\alpha_R^1V$ (non-linearities appear at larger voltages). This linear dependence in semi-conducting 2DEG is associated with the modification of the carrier density as a function of the bias \cite{nitta}. In the case of interfacial spin-orbit coupling, it is attributed to the variation of the orbital filling at the interface \cite{tsymbal}. Since the perpendicular anisotropy is quadratic in $\alpha_R$, we expect a linear dependence on the bias voltage: $T_{\bot}\propto\alpha_R^2\approx(\alpha_R^{02}+2\alpha_R^0\alpha_R^1V)$.\par

\subsection{Thin Barriers: Current-driven In-Plane Torque}
Let us now consider a thin barrier, similar to the ones used in regular tunneling spin torque studies \cite{review}, $d$=0.6 nm. When applying a bias voltage across the junction, spin polarized electrons flow through the tunnel barrier and their spin is reoriented due to interfacial SOI, which results in an out-of-equilibrium spin torque \cite{manchon2}. In the case of thin barriers, one expect that most of the voltage drop occurs within the barrier, therefore little modification of the Rashba coefficient is expected.\par

In this case, we find that besides the equilibrium perpendicular torque described above, a current-driven {\em in-plane} torque, $T_{||}$, appears. This in-plane torque displays a linear dependence on the bias voltage, as shown in Fig. 4(a). Interestingly, the out-of-equilibrium perpendicular torque is about one order of magnitude smaller than the in-plane torque [see inset in Fig. 4(b)]. The non-equilibrium in-plane torque rapidly vanishes when increasing the barrier height and thickness [Fig. 4(b)].\par

\begin{figure}[h!]
	\centering
		\includegraphics[width=8cm]{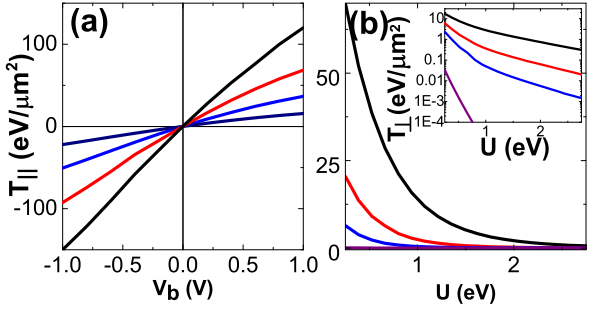}
	\caption{\label{fig:Fig4}(Color online) (a) Influence of bias voltage on the non-equilibrium in-plane torque $T_{||}$ for $\alpha_R\in$[2,5]eV.\AA$^2$; (b) Influence of barrier height $U$ on the non-equilibrium in-plane torque for $d\in[0.6,3]$nm. Inset: Influence of barrier height $U$ on the non-equilibrium perpendicular torque. In (b), $\alpha_R=5$eV.\AA$^2$ and $V_b=0.1$V.}
\end{figure}

This is an important result since the nature of the spin torque itself changes when modifying the barrier thickness: the non-equilibrium in-plane torque only appears in the limit of thin and low barriers, where equilibrium perpendicular torque is always present, giving rise to perpendicular magnetic anisotropy. 

\subsection{Comparison with STT and SOI-ST}

The present spin torque (referred to as TAMR-ST) can be readily compared with the conventional STT \cite{slonc} and the SOI-ST \cite{manchon}, as summarized in Table 1. On the first hand, it originates directly from the interfacial SOI rather than from the inhomogeneous magnetic texture. Therefore, it does not need any external polarizer. On the other hand, it possesses two components (like STT in MTJs), $T_{||}$ and $T_{\bot}$, whereas the SOI-ST produces only an effective magnetic field \cite{manchon,soitorque}. Furthermore, the in-plane component acts like a (anti-)damping ($\propto{\bf M}\times({\bf z}\times {\bf M})$) while the perpendicular component ($\propto{\bf z}\times {\bf M}$) competes with the demagnetizing field. Finally, the angular dependence of the torque in $\sin2\theta$ (versus $\sin\theta$ in MTJs) is expected to produce original current-driven magnetization dynamics \cite{manchon2}.

\begin{table}
	\centering
		\begin{tabular}{c|ccc}
			{\bf Torque} & STT & SOI-STT & TAMR-ST\\\hline\hline
			Origin & Magnetic texture & SOI & SOI\\
			Effective field & Yes in MTJs & Yes & Yes, perpendicular\\
			(anti-)damping & Yes & No & Yes\\
			Switching & Yes & Yes & Yes\\
			Precessions & Yes & No & Yes
		\end{tabular}\caption{Comparison between STT, SOI-ST and TAMR-Torque.}
\end{table}

\section{Device Implications}
\begin{figure}[h!]
	\centering
		\includegraphics[width=8cm]{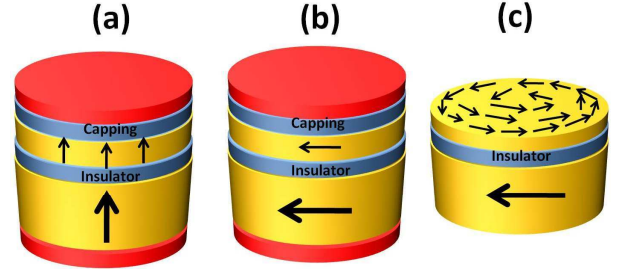}
	\caption{\label{fig:Fig5}Experimental set-ups: (a) perpendicular magnetic tunnel junction, (b) in-plane magnetic tunnel junction and (c) vortex core.}
\end{figure}

In this section, we discuss the possible implications of such a torque in the magnetization dynamics of three different devices based on conventional MTJ structures, comprising a pinned layer (bottom) and a free layer (top), separated by a thin insulating spacer (generally MgO). An additional capping layer, which can be either made of heavy metal or very thin insulator (thinner than the insulating spacer), can be inserted on the top of the free layer to ensure a significant TAMR-ST. These devices are depicted in Fig. \ref{fig:Fig5}. In this conventional MTJs, the free layer is submitted to a conventional spin torque on the form:
\begin{eqnarray}
{\bf T}_{mtj}=T_{||}^{mtj}{\bf M}\times({\bf P}\times{\bf M})+T_\bot^{mtj}{\bf P}\times{\bf M}
\end{eqnarray}
where ${\bf P}$ is the direction of the polarizer. The presence of interfacial SOI at both interfaces of the free layer generates an additional TAMR-ST:
\begin{eqnarray}
{\bf T}_{soi}=T_{||}^{soi}\cos\theta{\bf M}\times({\bf z}\times{\bf M})+T_\bot^{soi}\cos\theta{\bf z}\times{\bf M}\label{eq:eq2}
\end{eqnarray}
In Eq. (\ref{eq:eq2}), we explicitly accounted for the peculiar angular dependence of the torque in $\sin2\theta$. Since the perpendicular torque is essentially an equilibrium torque, as discussed above, its major influence is to create PMA. Most interestingly, the non-equilibrium in-plane torque $T_{||}^{soi}$ is linear in bias voltage, similarly to $T_{||}^{mtj}$, and can be in principle of the same order of magnitude.\par
The simplest configuration is given in Fig. \ref{fig:Fig5}(a), where the magnetization of both pinned and free layers are perpendicular to the plane (${\bf P}={\bf z}$). Then, both in-plane torque will compete against or add to each other, depending on the magnetic configure: parallel ($T_{||}^{mtj}+T_{||}^{soi}$), antiparallel ($T_{||}^{mtj}-T_{||}^{soi}$) or perpendicular ($T_{||}^{mtj}$) magnetic configuration. The dynamics is expected to be much more complex in in-plane MTJs [Fig. 5(b)], since ${\bf P}\neq{\bf z}$. Although the TAMR-ST will not participate to the onset of the magnetization switching ($\cos\theta\approx0$ for $\theta\approx\pi/2$), it will have a strong influence on the magnetization dynamics itself, favoring either out-of-plane or in-plane states depending on the sign of the bias voltage.
Finally, Fig. \ref{fig:Fig5}(c) shows a configuration similar to the one used in Ref. \cite{fert}, where the spin torque acts on a vortex wall. In the absence of TAMR-ST, the in-plane STT excites strong vortex oscillations \cite{fert}. In presence of interfacial SOI, the in-plane TAMR-ST is expected to dramatically modify the magnetic texture: Since this torque tends to align the magnetization either in the plane or out of the plane of the layers, it is expected to either broaden or narrow the vortex core, depending on the bias voltage direction. More detailed will be given elsewhere.

\section{Conclusion}
Spin transport in a SMTJ in the presence of interfacial Rashba SOI was investigated theoretically. Besides the well-known TAMR effect, we showed that a spin torque arises from the presence of interfacial SOI. In the large and thick barrier limit, the conductance vanishes and the equilibrium torque induces PMA that can be in principle controlled by interfacial SOI. In the opposite limit of low and thin barriers (large conductance), a non-equilibrium torque arises that allows for magnetization excitations.

\section*{Acknowledgments}
The author aknowledges fruitful discussions with S. Zhang, K.-J. Lee and P. J. Kelly. This work was supported by KAUST AEA funding program.

\end{document}